\documentclass[journal]{IEEEtran}
\usepackage{amsmath}
\usepackage{graphicx}
\usepackage{color}
\usepackage{ulem}
\usepackage{amssymb}

\begin{document}
\title{Convex Optimization over Fixed Value Point Set of Quasi-Nonexpansive Random Operators on Hilbert Spaces}
\author{S. Sh. Alaviani
\thanks{ }
\thanks{S. Sh. Alaviani is with the Department of Mechanical Engineering, Clemson University, Clemson,
	SC, 29634 USA e-mail: salavia@clemson.edu. }
}

\maketitle

\begin{abstract}
In this paper, a new optimization framework is defined that includes the optimization framework recently proposed in \cite{alavianiACC17}-\cite{alavianiTAC} as a special case. The convex optimization in \cite{alavianiACC17}-\cite{alavianiTAC} includes centralized optimization and distributed optimization over random networks, so does the optimization defined here. It is shown that the proposed algorithm in \cite{alavianiTAC} converges almost surely and in mean square to  a solution of the optimization problem here under suitable assumptions. 
\end{abstract}


\section{Introduction}

Optimization has applications in many areas of engineering such as signal recovery \cite{signal} and data regression \cite{dataregression}. Lagrangian framework has been a useful tool, by adding dual variables, to elucidate centralized \cite{boyd} and distributed optimization over non-switching networks \cite{nicola1}-\cite{nicola2}. Also, Game Theory has been helpful to explain distributed optimization over non-switching networks \cite{shamma}-\cite{nali}. Nevertheless, the aforementioned methods cannot interpret the nature of distributed optimization over random networks.

Recently, by defining a new mathematical terminology called \textit{fixed value point}, an optimization framework has been developed in \cite{alavianiACC17}-\cite{alavianiTAC}, i.e., minimization of a convex function over fixed value point set of nonexpansive random operators on Hilbert spaces. The optimization includes centralized and distributed optimization over random networks (see \cite{alavianiCDC}-\cite{alavianidissertation} for more details). An algorithm has been proposed in \cite{alavianiTAC} to converge almost surely and in mean square to a solution of the problem under suitable assumptions.

In this paper, we define a new optimization framework, namely minimization of a convex function over fixed value point set of a quasi-nonexpansive random operators on Hilbert spaces. This optimization includes the optimization framework defined in \cite{alavianiACC17}-\cite{alavianiTAC} as a special case. Therefore, this optimization includes centralized optimization and distributed optimization over random networks. We show that the proposed algorithm in \cite{alavianiTAC} can converge almost surely and in mean square to the global solution of this optimization problem under suitable assumptions.

This paper is organized as follows. In Section II, preliminaries on random operators and stochastic convergence are provided. The mathematical optimization problem and the discrete-time algorithm for almost sure and in mean square convergence to the solution are in Section III.

\textit{Notations:} $\Re$ denotes the set of all real numbers. $\Re^{n}$ denotes the $n$-dimensional real space. $\mathbb{N}$ denotes the set of all natural numbers. $\times$ represents Cartesian product. $E[x]$ denotes Expectation of the random variable $x$. $\emptyset$ denotes the empty set. $\nabla f(x)$ represents the gradient of $f$ at $x$.

\section{Preliminaries}

Let $\mathcal{H}$ be a real Hilbert space with norm $\Vert . \Vert $ and inner product $\langle .,. \rangle$. An operator $A:\mathcal{H} \longrightarrow \mathcal{H}$ is said to be \textit{monotone} if $\langle x-y,Ax-Ay \rangle  \geq 0$ for all $x,y \in \mathcal{H}$. $A:\mathcal{H} \longrightarrow \mathcal{H}$ is called $\rho$\textit{-strongly monotone} if $\langle x-y,Ax-Ay \rangle \geq \rho \Vert x-y \Vert^{2}$ for all $x,y \in \mathcal{H}$. A function $f(x)$ is $\rho$\textit{-strongly convex} if $\langle x-y,\nabla f(x)-\nabla f(y) \rangle \geq \rho \Vert x-y \Vert^{2}$ for all $x,y \in \mathcal{H}$. Therefore, a function is $\rho$-strongly convex if its gradient is $\rho$-strongly monotone. 

A mapping $B:\mathcal{H} \longrightarrow \mathcal{H}$ is said to be $K$\textit{-Lipschitz continuous} if there exists a $K > 0$ such that $\Vert Bx-By \Vert \leq K \Vert x-y \Vert$ for all $x,y \in \mathcal{H}$. Let $S$ be a nonempty subset of a Hilbert space $\mathcal{H}$ and $Q:S \longrightarrow \mathcal{H}$. The point $x$ is called a \textit{fixed point} of $Q$ if $x=Q(x)$. Let $\omega^{*}$ and $\omega$ denote elements in the sets $\Omega^{*}$ and $\Omega$, respectively.

Let $(\Omega^{*},\sigma)$ be a measurable space ($\sigma$-sigma algebra) and $C$ be a nonempty subset of a Hilbert space $\mathcal{H}$. A mapping $x:\Omega^{*} \longrightarrow \mathcal{H}$ is \textit{measurable} if $x^{-1}(U) \in \sigma$ for each open subset $U$ of $\mathcal{H}$. The mapping $T:\Omega^{*} \times C \longrightarrow \mathcal{H}$ is a \textit{random map} if for each fixed $z \in C$, the mapping $T(.,z):\Omega^{*} \longrightarrow \mathcal{H}$ is measurable, and it is \textit{continuous} if for each $\omega^{*} \in \Omega^{*}$ the mapping $T(\omega^{*},.):C \longrightarrow \mathcal{H}$ is continuous.

\textbf{Definition 1:} A measurable mapping $x:\Omega^{*} \longrightarrow \mathcal{H}$ is a \textit{random fixed point} of the random map $T:\Omega^{*} \times C \longrightarrow \mathcal{H}$ if $T(\omega^{*},x(\omega^{*}))=x(\omega^{*})$ for each $\omega^{*} \in \Omega^{*}$.

\textbf{Definition 2} \cite{alavianiACC17}-\cite{alavianiTAC}: If there exists a point $\hat{x} \in \mathcal{H}$ such that $\hat{x}=T(\omega^{*},\hat{x})$ for all $\omega^{*} \in \Omega^{*}$, it is called \textit{fixed value point}, and $FVP(T)$ represents the set of all fixed value points of $T$. 

\textbf{Remark 1} \cite{alavianiACC17}-\cite{alavianiTAC}: A random mapping may have a random fixed point but may not have a fixed value point. For instance, if $\Omega^{*}=\lbrace H,G \rbrace $ and $T(H,x(H))=1, T(G,x(G))=0$, then the random variable $x(H)=1,x(G)=0$ is a random fixed point of $T$. However, $T$ does not have any fixed value point. 

\textbf{Definition 3:} Let $C$ be a nonempty subset of a Hilbert space $\mathcal{H}$ and $T:\Omega^{*} \times C \longrightarrow C$ be a random map. The map $T$ is said to be \textit{nonexpansive random operator} if for each $\omega^{*} \in \Omega^{*}$ and for arbitrary $x,y \in C$ we have 
$$\Vert T(\omega^{*} ,x)-T(\omega^{*} ,y) \Vert \leq \Vert x-y \Vert.$$

\textbf{Definition 4:} Let $C$ be a nonempty subset of a Hilbert space $\mathcal{H}$ and $T:\Omega^{*} \times C \longrightarrow C$ be a random map. The map $T$ is said to be \textit{quasi-nonexpansive random operator} if for any $x \in C$ we have
$$\Vert T(\omega^{*},x)-\xi(\omega^{*}) \Vert \leq \Vert x-\xi(\omega^{*}) \Vert$$
where $\xi:\Omega^{*} \longrightarrow C$ is a random fixed point of $T$.

\textbf{Remark 2:} Every nonexpansive random operator is a quasi-nonexpansive random operator if it has a random fixed point.

\textbf{Remark 3:} If a quasi-nonexpansive random operator has a fixed value point, namely $x^{*}$, then we have, from Definitions 2 and 4, for any $x \in C$ that
$$\Vert T(\omega^{*},x)-x^{*} \Vert \leq \Vert x-x^{*} \Vert.$$

The following preposition is a corollary of Theorem 1 in \cite{quasinonexpansive}.

\textbf{Preposition 1:} If $C$ is a closed convex subset of a Hilbert space $\mathcal{H}$, and $T:C \longrightarrow C$ is quasi-nonexpansive, then $Fix(T)$ is a nonempty closed convex set.

\textbf{Definition 5:} A sequence of random variables $x_{n}$ is said to \textit{converge pointwise (surely)} to $x$ if for every $\omega \in \Omega$
$$\lim_{n \longrightarrow \infty} \Vert x_{n}(\omega)-x(\omega) \Vert=0.$$

\textbf{Definition 6:} A sequence of random variables $x_{n}$ is said to \textit{converge almost surely} to $x$ if there exists a subset $A \subseteq \Omega$ such that $Pr(A)=0$, and for every $\omega \notin A$
$$\lim_{n \longrightarrow \infty} \Vert x_{n}(\omega)-x(\omega) \Vert=0.$$

\textbf{Definition 7:} A sequence of random variables $x_{n}$ is said to \textit{converge in mean square} to $x$ if 
$$E [\Vert x_{n}-x \Vert^{2}] \longrightarrow 0 \quad{as} \quad{ n \longrightarrow \infty}.$$

\textbf{Lemma 1} \cite{ssdd}: Let $\lbrace a_{n}  \rbrace_{n=0}^{\infty}$ be a sequence of non-negative real numbers satisfying
$$a_{n+1} \leq (1-b_{n})a_{n}+b_{n}h_{n}+c_{n}$$ 
where $b_{n} \in [0,1], \sum_{n=0}^{\infty} b_{n}=\infty,\displaystyle \limsup_{n \longrightarrow \infty} h_{n} \leq 0,$ and $\sum_{n=0}^{\infty} c_{n} < \infty$. Then 
$$\lim_{n \longrightarrow \infty} a_{n}=0.$$

\textbf{Lemma 2:} Let the sequence $\{ x_{n} \}_{n=0}^{\infty}$ in a real Hilbert space $\mathcal{H}$ be bounded for each realization $\omega \in \Omega$ and converge almost surely to $x^{*}$. Then the sequence converges in mean square to $x^{*}.$

\textit{Proof:} See the proof of Theorem 2 in \cite{alavianiTAC}.

\section{Main Results}

Now we define a new optimization problem stated in Problem 1 below.

\textbf{Problem 1:} Let $\mathcal{H}$ be a real Hilbert space. Assume that the problem is feasible, namely $FVP(T) \neq \emptyset$. Given a convex function $f:\mathcal{H} \longrightarrow \Re$ and a quasi-nonexpansive random mapping $T:\Omega^{*} \times \mathcal{H} \longrightarrow \mathcal{H}$, the problem is to find $x^{*} \in \underset{x}{\operatorname{argmin}} f(x)$  such that $x^{*} $ is a fixed value point of $T(\omega^{*},x)$, i.e., we have the following minimization problem

\begin{equation}\label{prob1}
\begin{aligned}
& \underset{x}{\text{min}}
& & f(x) \\
& \text{subject to}
& & x \in FVP(T)
\end{aligned}
\end{equation}
where $FVP(T)$ is the set of fixed value points of the random operator $T(\omega^{*},x)$ (see Definition 2).

\textbf{Remark 4:} A fixed value point of a quasi-nonexpansive random mapping is a common fixed point of a family of quasi-nonexpansive non-random mappings $T(\omega^{*},.)$ for each $\omega^{*}$. From Preposition 1, the fixed point set of a quasi-nonexpansive non-random mapping $T(\omega^{*},.)$ for each $\omega^{*}$ is a convex set. It is known that the intersection of convex sets (finite, countable, or uncountable) is convex. Therefore, $FVP(T)$ is a convex set, and Problem 1 is a convex optimization problem.

\textbf{Remark 5:} According to Remarks 2 and 3, Problem 1 includes Problem 3 in \cite{alavianiTAC} as a special case.

Based on the work of \cite{alavianiTAC}, we propose the following algorithm for solving Problem 1:
\begin{equation}\label{7}
x_{n+1}=\alpha_{n} (x_{n}- \beta  \nabla f(x_{n}))+(1-\alpha_{n}) \hat{T}(\omega_{n}^{*},x_{n}),
\end{equation}
where $\hat{T}(\omega_{n}^{*},x_{n})=(1-\eta) x_{n}+\eta T(\omega_{n}^{*},x_{n}), \eta \in (0,1), \alpha_{n} \in [0,1].$

Let $(\Omega^{*}, \sigma)$ be a measurable space where $\Omega^{*}$ and $\sigma$ are defined in Section II. Consider a probability measure $\mu$ defined on the space $(\Omega,\mathcal{F})$ where
\begin{align*}
\Omega &=\Omega^{*} \times \Omega^{*} \times \Omega^{*} \times \hdots\\
\mathcal{F} &=\sigma \times \sigma \times \sigma \times \hdots
\end{align*}
such that $(\Omega,\mathcal{F},\mu)$ forms a probability space. We denote a realization in this probabilty space by $\omega \in \Omega$.

Now we impose the following assumptions.

\textbf{Assumption 1:} $f(x)$ is $\rho$-strongly convex, and $\nabla f(x)$ is $K$-Lipschitz continuous.

\textbf{Assumption 2:} There exists a nonempty subset $\tilde{K} \subseteq \Omega^{*}$ such that $FVP(T)=\{ \tilde{z} | \tilde{z} \in \mathcal{H}, \tilde{z}=T(\bar{\omega},\tilde{z}), \forall \bar{\omega} \in \tilde{K} \}$, and each element of $\tilde{K}$ occurs infinitely often almost surely.

Before we give our theorems, we need to extend Lemma 5 in \cite{alavianiTAC} for quasi-nonexpansive random operators. Hence, we have the following lemma.

\textbf{Lemma 3:} Let $\mathcal{H}$ be a real Hilbert space, $\hat{T}(\omega^{*},x):=(1-\eta)x+\eta T(\omega^{*},x), \omega^{*} \in \Omega^{*}, x \in \mathcal{H},$ with a quasi-nonexpansive random operator $T$, $FVP(T) \neq \emptyset$, and $\eta \in (0,1]$. Then

\textit{(i)} $FVP(T)=FVP(\hat{T}).$

\textit{(ii)} $\langle x-\hat{T}(\omega^{*},x),x-z \rangle \geq \frac{\eta}{2} \Vert x-T(\omega^{*},x) \Vert^{2}, \forall z \in \quad{} \quad{} \quad{} \quad{} FVP(T), \forall \omega^{*} \in \Omega^{*}.$

\textit{(iii)} $\hat{T}(\omega^{*},x)$ is quasi-nonexpasnive.

\textit{Proof:} See Appendix I.

Now we give the main theorem in this paper.

\textbf{Theorem 1:} Consider Problem 1 with Assumptions 1 and 2. Let $\beta \in (0,\frac{2 \rho}{K^{2}})$ and $\alpha_{n} \in [0,1], n \in \mathbb{N} \cup \lbrace 0 \rbrace$ such that

\textit{(a)} $\displaystyle \lim_{n \longrightarrow \infty} \alpha_{n}=0,$

\textit{(b)} $\sum_{n=0}^{\infty} \alpha_{n}=\infty.$

Then starting from any initial point, the sequence generated by (\ref{7}) globally converges almost surely to the unique solution of the problem.

\textbf{Remark 6:} An example of $\alpha_{n}$ satisfying \textit{(a)} and \textit{(b)} is $\alpha_{n} :=\frac{1}{(1+n)^{\zeta}}$ where $\zeta \in (0,1]$.

\textit{Proof of Theorem 1:} 

We prove Theorem 1 in three steps.

\textit{Step 1: $\{ x_{n} \}_{n=0}^{\infty}, \forall \omega \in \Omega,$ is bounded.}

Since the cost function is strongly convex and the constraint set is closed, the problem has a unique solution. Let $x^{*}$ be the unique solution of the problem. We can write $x^{*}=\alpha_{n} x^{*} +(1-\alpha_{n})x^{*}, \forall n \in \mathbb{N} \cup \{ 0 \}$. Therefore, we have
\begin{align*}
\Vert x_{n+1}-x^{*} \Vert &=\Vert \alpha_{n} (x_{n}- \beta \nabla f(x_{n}))  \\
&\quad +(1-\alpha_{n}) \hat{T}(\omega_{n}^{*},x_{n})-x^{*} \Vert \\
&= \Vert \alpha_{n}(x_{n}-\beta \nabla f(x_{n})-x^{*}) \\
&\quad +(1-\alpha_{n})(\hat{T}(\omega_{n}^{*},x_{n})-x^{*}) \Vert \\
&\leq \alpha_{n}  \Vert x_{n}-\beta \nabla f(x_{n})-x^{*} \Vert \\
& \quad +(1-\alpha_{n})  \Vert \hat{T}(\omega_{n}^{*},x_{n})-x^{*} \Vert.
\end{align*}
Since $x^{*}$ is the solution, we have that $x^{*} \in FVP(T)=FVP(\hat{T})$ (see part \textit{(i)} of Lemma 3). Due to the fact that $\hat{T}(\omega^{*},x)$ is a quasi-nonexpansive random operator (see part \textit{(iii)} of Lemma 3), the above can be written as
\begin{align}
\Vert x_{n+1}-x^{*} \Vert &\leq \alpha_{n}  \Vert x_{n}-\beta \nabla f(x_{n})-x^{*} \Vert \nonumber \\
&\quad +(1-\alpha_{n})  \Vert \hat{T}(\omega_{n}^{*},x_{n})-x^{*} \Vert \nonumber \\
& \leq \alpha_{n}  \Vert x_{n}-\beta \nabla f(x_{n})-x^{*} \Vert \nonumber \\
&\quad +(1-\alpha_{n})  \Vert x_{n}-x^{*} \Vert. \label{9}
\end{align}
In a Hilbert space $\mathcal{H}$, we have
\begin{equation}\label{hilbertinequ}
\Vert u+v \Vert^{2}=\Vert u \Vert^{2}+\Vert v \Vert^{2}+2 \langle u,v \rangle, \forall u,v \in \mathcal{H}.
\end{equation}
Since $\nabla f(x)$ is $\rho$-strongly monotone, and $\nabla f(x)$ is $K$-Lipschitz continuous, we obtain from (\ref{hilbertinequ}) for any $x,y \in \mathcal{H}$ that
\begin{align*}
&\quad \Vert x-y-\beta (\nabla f(x)-\nabla f(y)) \Vert^{2} \\
&=\Vert x-y \Vert^{2}  \\
&\quad -2 \beta \langle \nabla f(x)-\nabla f(y),x-y \rangle \\
&\quad +\beta^{2} \Vert \nabla f(x)-\nabla f(y) \Vert^{2} \\
&\leq \Vert x-y \Vert^{2} \\
&\quad -2 \rho \beta \Vert x-y \Vert^{2} \quad{} \text{(strong convexity)}\\
&\quad +K^{2} \beta^{2} \Vert x-y \Vert^{2} \quad{} \text{($K$-Lipschitz)} \\
&= (1-2 \rho \beta+\beta^{2} K^{2}) \Vert x-y \Vert^{2} \\
&= (1-\gamma)^{2} \Vert x-y \Vert^{2}
\end{align*}
where $\gamma:=1-\sqrt{1-\beta(2 \rho-\beta K^{2})}$, and selecting $\beta \in (0,\frac{2 \rho}{K^{2}})$ implies $0< \gamma \leq 1$. Indeed, we have
\begin{equation}\label{10}
\Vert x-y-\beta(\nabla f(x)-\nabla f(y)) \Vert \leq (1-\gamma) \Vert x-y \Vert. 
\end{equation}
We have that
\begin{align}
&\quad \Vert x_{n}-\beta \nabla f(x_{n})-x^{*} \Vert   \nonumber \\
&= \Vert x_{n}-x^{*}-\beta(\nabla f(x_{n})-\nabla f(x^{*}))-\beta \nabla f(x^{*}) \Vert  \nonumber \\
&\leq \Vert x_{n}-x^{*}-\beta(\nabla f(x_{n})-\nabla f(x^{*})) \Vert +\beta  \Vert \nabla f(x^{*}) \Vert. \label{beta11} 
\end{align}
Therefore, (\ref{10}) and (\ref{beta11}) implies
\begin{align}
\Vert x_{n}-\beta \nabla f(x_{n})-x^{*} \Vert &\leq \Vert x_{n}-x^{*}-\beta(\nabla f(x_{n})-\nabla f(x^{*})) \Vert \nonumber \\
&\quad +\beta  \Vert \nabla f(x^{*}) \Vert \nonumber \\
&\leq (1-\gamma)  \Vert x_{n}-x^{*} \Vert \nonumber \\
&\quad + \beta  \Vert \nabla f(x^{*}) \Vert. \label{11}
\end{align}
Substituting (\ref{11}) for (\ref{9}) yields
\begin{align*}
\Vert x_{n+1}-x^{*} \Vert &\leq (1-\gamma \alpha_{n})  \Vert x_{n}-x^{*} \Vert \\
&\quad +\alpha_{n} \beta \Vert \nabla f(x^{*}) \Vert \\
&= (1-\gamma \alpha_{n})  \Vert x_{n}-x^{*} \Vert \\
&\quad + \gamma \alpha_{n}(\frac{\beta \Vert \nabla f(x^{*}) \Vert}{\gamma})
\end{align*}
which by induction implies that
$$ \Vert x_{n+1}-x^{*} \Vert \leq max \lbrace  \Vert x_{0}-x^{*} \Vert, \frac{\beta \Vert \nabla f(x^{*}) \Vert}{\gamma}  \rbrace$$
that implies $ \Vert x_{n}-x^{*} \Vert, n \in \mathbb{N} \cup \lbrace 0 \rbrace, \forall \omega \in \Omega$, is bounded. Therefore, $\{ x_{n} \}_{n=0}^{\infty}$ is bounded for all $\omega \in \Omega$.

\textit{Step 2: $\{ x_{n} \}_{n=0}^{\infty}$ converges almost surely to a random variable supported by the feasible set.}

From (\ref{7}) and $x_{n}=\alpha_{n} x_{n}+(1-\alpha_{n}) x_{n}$, we have
\begin{equation}\label{march0}
x_{n+1}-x_{n}+\alpha_{n} \beta  \nabla f(x_{n})=(1-\alpha_{n}) (\hat{T}(\omega_{n}^{*},x_{n})-x_{n}),
\end{equation}
and hence
\begin{align}
&\quad \langle x_{n+1}-x_{n}+\alpha_{n} \beta  \nabla f(x_{n}),x_{n}-x^{*} \rangle  \nonumber \\
&=-(1-\alpha_{n})  \langle x_{n}-\hat{T}(\omega_{n}^{*},x_{n}),x_{n}-x^{*} \rangle. \label{march1}
\end{align}
Since $x^{*} \in FVP(T)$, we have from part \textit{(ii)} of Lemma 3 that
\begin{equation}\label{march2}
\langle x_{n}-\hat{T}(\omega_{n}^{*},x_{n}),x_{n}-x^{*} \rangle \geq \frac{\eta}{2} \Vert x_{n}-T(\omega_{n}^{*},x_{n}) \Vert^{2}.
\end{equation}
From (\ref{march1}) and (\ref{march2}), we obtain
\begin{align}
&\quad \langle x_{n+1}-x_{n}+\alpha_{n} \beta  \nabla f(x_{n}),x_{n}-x^{*} \rangle  \nonumber \\
&\leq - \frac{\eta}{2} (1-\alpha_{n}) \Vert x_{n}-T(\omega_{n}^{*},x_{n}) \Vert^{2} \label{march3}
\end{align}
or equivalently
\begin{align}
&\quad- \langle x_{n}-x_{n+1},x_{n}-x^{*} \rangle \nonumber \\
&\leq -\alpha_{n} \langle \beta \nabla f(x_{n}),x_{n}-x^{*} \rangle  \nonumber \\
&\quad -\frac{\eta}{2} (1-\alpha_{n}) \Vert x_{n}-T(\omega_{n}^{*},x_{n}) \Vert^{2}. \label{march4}
\end{align}
For any $u,v \in \mathcal{H}$ we have
\begin{equation}\label{march5}
\langle u,v \rangle =-\frac{1}{2} \Vert u-v \Vert^{2}+\frac{1}{2} \Vert u \Vert^{2}+\frac{1}{2} \Vert v \Vert^{2}.
\end{equation}
From (\ref{march5}) we obtain
\begin{equation}\label{march6}
\langle x_{n}-x_{n+1},x_{n}-x^{*}\rangle=-C_{n+1}+C_{n}+\frac{1}{2} \Vert x_{n}-x_{n+1} \Vert^{2}
\end{equation}
where $C_{n}:=\frac{1}{2} \Vert x_{n}-x^{*} \Vert^{2}$. From (\ref{march4}) and (\ref{march6}) we obtain
\begin{align}
&\quad C_{n+1}-C_{n}-\frac{1}{2} \Vert x_{n}-x_{n+1} \Vert^{2} \nonumber \\
&\leq - \alpha_{n} \langle \beta \nabla f(x_{n}),x_{n}-x^{*} \rangle \nonumber \\
&\quad -\frac{\eta}{2} (1-\alpha_{n}) \Vert x_{n}-T(\omega_{n}^{*},x_{n}) \Vert^{2}. \label{march7} 
\end{align}
From (\ref{march0}) and (\ref{hilbertinequ}) we have
\begin{align}
&\quad \Vert x_{n+1}-x_{n} \Vert^{2}   \nonumber \\
&= \Vert - \alpha_{n} \beta \nabla f(x_{n})+(1-\alpha_{n}) (\hat{T}(\omega^{*}_{n},x_{n})-x_{n}) \Vert^{2}  \nonumber \\
&= \alpha_{n}^{2}  \Vert \beta \nabla f(x_{n}) \Vert^{2}+ (1-\alpha_{n})^{2} \Vert \hat{T}(\omega^{*}_{n},x_{n})-x_{n} \Vert^{2} \nonumber\\
&\quad - 2 \alpha_{n} (1-\alpha_{n}) \langle \beta \nabla f(x_{n}),\hat{T}(\omega_{n}^{*},x_{n})-x_{n} \rangle.  \label{march9} 
\end{align}
We know that $\Vert \hat{T}(\omega^{*}_{n},x_{n})-x_{n} \Vert=\eta \Vert x_{n}-T(\omega^{*}_{n},x_{n}) \Vert$. Since $\alpha_{n} \in [0,1]$, we have also that $(1-\alpha_{n})^{2} \leq (1-\alpha_{n})$. Using these facts as well as multiplying both sides of (\ref{march9}) by $\frac{1}{2}$ yield
\begin{align}
&\quad \frac{1}{2} \Vert x_{n+1}-x_{n} \Vert^{2}  \nonumber \\
&= \frac{1}{2} \alpha_{n}^{2}  \Vert \beta \nabla f(x_{n}) \Vert^{2}   \nonumber \\
&\quad +\frac{1}{2} (1-\alpha_{n})^{2} \eta^{2} \Vert T(\omega^{*}_{n},x_{n})-x_{n} \Vert^{2} \nonumber \\
&\quad - \alpha_{n} (1-\alpha_{n}) \langle \beta \nabla f(x_{n}),\hat{T}(\omega_{n}^{*},x_{n})-x_{n} \rangle.  \nonumber \\
&\leq \frac{1}{2} \alpha_{n}^{2}  \Vert \beta \nabla f(x_{n}) \Vert^{2}   \nonumber \\
&\quad +\frac{1}{2} (1-\alpha_{n}) \eta^{2} \Vert T(\omega^{*}_{n},x_{n})-x_{n} \Vert^{2} \nonumber \\
&\quad - \alpha_{n} (1-\alpha_{n}) \langle \beta \nabla f(x_{n}),\hat{T}(\omega_{n}^{*},x_{n})-x_{n} \rangle.  \label{march10}
\end{align}
From (\ref{march7}) and (\ref{march10}), we obtain
\begin{align}
C_{n+1}-C_{n} &\leq \frac{1}{2} \Vert x_{n+1}-x_{n} \Vert^{2} \nonumber\\
&\quad - \alpha_{n} \langle \beta \nabla f(x_{n}),x_{n}-x^{*} \rangle \nonumber\\
&\quad -\frac{\eta}{2} (1-\alpha_{n}) \Vert x_{n}-T(\omega_{n}^{*},x_{n}) \Vert^{2} \nonumber\\
& \leq -(\frac{1}{2}-\frac{\eta}{2}) \eta  (1-\alpha_{n}) \Vert x_{n}-T(\omega_{n}^{*},x_{n}) \Vert^{2} \nonumber\\
&\quad + \alpha_{n}(\frac{1}{2}\alpha_{n} \Vert \beta \nabla f(x_{n}) \Vert^{2} \nonumber\\
&\quad -\langle \beta \nabla f(x_{n}),x_{n}-x^{*} \rangle \nonumber\\
&\quad -(1-\alpha_{n}) \langle \beta \nabla f(x_{n}),\hat{T}(\omega_{n}^{*},x_{n})-x_{n} \rangle). \label{march11}  
\end{align}
Now we claim that there exists an $n_{0} \in \mathbb{N}$ such that the sequence $\{ C_{n} \}$ is non-increasing for $n \geq n_{0}$. Assume by contradiction that this is not true. Then there exists a subsequence $\{ C_{n_{j}} \}$ such that
$$C_{n_{j}+1}-C_{n_{j}}>0$$
which together with (\ref{march11}) yields
\begin{align}
0 &<C_{n_{j}+1}-C_{n_{j}} \nonumber\\
&\leq -(\frac{1}{2}-\frac{\eta}{2}) \eta  (1-\alpha_{n_{j}}) \Vert x_{n_{j}}-T(\omega_{n_{j}}^{*},x_{n_{j}}) \Vert^{2} \nonumber\\
&\quad + \alpha_{n_{j}}(\frac{1}{2} \alpha_{n_{j}} \beta^{2}\Vert \nabla f(x_{n_{j}}) \Vert^{2} \nonumber\\
&\quad -\langle \beta \nabla f(x_{n_{j}}),x_{n_{j}}-x^{*} \rangle \nonumber\\
&\quad -(1-\alpha_{n_{j}}) \langle \beta \nabla f(x_{n_{j}}),\hat{T}(\omega_{n_{j}}^{*},x_{n_{j}})-x_{n_{j}} \rangle). \label{march12}    
\end{align}
Since $ \{ x_{n} \}$ is bounded, $\nabla f(x)$ is continuous, and $\eta \in (0,1)$,  we obtain from (\ref{march12}) by Theorem 1 \textit{(a)} that
\begin{align}
0 &< \liminf_{j \longrightarrow \infty} [-(\frac{1}{2}-\frac{\eta}{2}) \eta  (1-\alpha_{n_{j}}) \Vert x_{n_{j}}-T(\omega_{n_{j}}^{*},x_{n_{j}}) \Vert^{2} \nonumber \\
&\quad +\alpha_{n_{j}}(\frac{1}{2} \alpha_{n_{j}} \Vert \beta \nabla f(x_{n_{j}}) \Vert^{2} \nonumber \\
&\quad -\langle \beta \nabla f(x_{n_{j}}),x_{n_{j}}-x^{*} \rangle \nonumber \\
&\quad -(1-\alpha_{n_{j}}) \langle \beta \nabla f(x_{n_{j}}),\hat{T}(\omega_{n_{j}}^{*},x_{n_{j}})-x_{n_{j}} \rangle)] \nonumber \\
& \leq 0
\end{align}
which is a contradiction. Therefore, there exists an $n_{0} \in \mathbb{N}$ such that the sequence $\{ C_{n} \}$ is non-increasing for $n \geq n_{0}$. Since $\{ C_{n} \}$ is bounded below, it converges for all $\omega \in \Omega$.

Taking the limit of both sides of (\ref{march11}) and using the convergence of $\{ C_{n} \}$, continuity of $\nabla f(x)$, \textit{Step 1}, $\eta \in (0,1)$, and Theorem 1 \textit{(a)} yield 
$$\lim_{n \longrightarrow \infty} \Vert x_{n}-T(\omega_{n}^{*},x_{n}) \Vert=0, \quad{} \textit{pointwise}$$  
which implies that $\{ x_{n} \}_{n=0}^{\infty}$ converges for each $\omega \in \Omega$ since $FVP(T) \neq \emptyset$. Moreover, this together with Assumption 2 implies that $\{ x_{n} \}$ converges almost surely to a random variable supported by $FVP(T)$.

\textit{Step 3: $\{ x_{n} \}_{n=0}^{\infty}$ converges almost surely to the optimal solution.}

It remains to prove that $\{ x_{n} \}_{n=0}^{\infty}$ converges almost surely to the optimal solution. Since $x^{*} \in FVP(T)$ is the optimal solution, we have
\begin{equation}\label{optimality}
\langle \bar{x}-x^{*},\nabla f(x^{*}) \rangle \geq 0, \forall \bar{x} \in FVP(T).
\end{equation}

We have from (\ref{hilbertinequ}) that 
\begin{align}
&\quad \Vert x_{n+1}-x^{*} \Vert^{2}   \nonumber \\
&= \Vert x_{n+1}-x^{*}+\alpha_{n} \beta \nabla f(x^{*})-\alpha_{n} \beta \nabla f(x^{*}) \Vert^{2} \nonumber \\
&= \Vert x_{n+1}-x^{*}+\alpha_{n} \beta \nabla f(x^{*}) \Vert^{2}+ \alpha_{n}^{2}  \Vert \beta \nabla f(x^{*}) \Vert^{2} \nonumber\\
&\quad -2 \alpha_{n}   \langle \beta \nabla f(x^{*}),x_{n+1}-x^{*}+\alpha_{n} \beta \nabla f(x^{*}) \rangle.  \label{19}
\end{align}
We have that $x^{*}=\alpha_{n}x^{*}+(1-\alpha_{n})x^{*}, \forall n \in \mathbb{N} \cup  \{ 0 \}$; using this fact and (\ref{7}), we obtain 
\begin{align}
&\quad \Vert x_{n+1}-x^{*}+\alpha_{n} \beta \nabla f(x^{*}) \Vert^{2}   \nonumber \\
&=\Vert \alpha_{n}[x_{n}-x^{*}-\beta( \nabla f(x_{n})- \nabla f(x^{*}))] \nonumber \\
&\quad +(1-\alpha_{n})[\hat{T}(\omega_{n}^{*},x_{n})-x^{*}] \Vert^{2}.  \label{20}
\end{align}
Furthermore, we have
\begin{align}
&\quad \langle \beta \nabla f(x^{*}),x_{n+1}-x^{*}+\alpha_{n} \beta \nabla f(x^{*}) \rangle  \nonumber \\
&= \langle \beta \nabla f(x^{*}),x_{n+1}-x^{*} \rangle \nonumber \\
&\quad +\alpha_{n} \langle \beta \nabla f(x^{*}),\beta \nabla f(x^{*}) \rangle \nonumber \\
&= \langle \beta \nabla f(x^{*}),x_{n+1}-x^{*} \rangle \nonumber \\
&\quad +\alpha_{n}  \Vert \beta \nabla f(x^{*}) \Vert^{2}. \label{21}
\end{align}
Substituting (\ref{20}) and (\ref{21}) for (\ref{19}) yields
\begin{align*}
&\quad \Vert x_{n+1}-x^{*} \Vert^{2} \nonumber\\
&=\Vert x_{n+1}-x^{*}+\alpha_{n} \beta \nabla f(x^{*}) \Vert^{2} \nonumber \\
&\quad + \alpha_{n}^{2}  \Vert \beta \nabla f(x^{*}) \Vert^{2} \nonumber\\
&\quad -2 \alpha_{n}  \langle \beta \nabla f(x^{*}),x_{n+1}-x^{*}+\alpha_{n} \beta \nabla f(x^{*}) \rangle \nonumber\\
&=\Vert \alpha_{n}[x_{n}-x^{*}-\beta(\nabla f(x_{n})-\nabla f(x^{*}))] \nonumber\\
&\quad +(1-\alpha_{n})[\hat{T}(\omega_{n}^{*},x_{n})-x^{*}] \Vert^{2} \nonumber\\
&\quad -2 \alpha_{n} \langle \beta \nabla f(x^{*}),x_{n+1}-x^{*} \rangle - \alpha_{n}^{2} \Vert \beta \nabla f(x^{*}) \Vert^{2} \nonumber\\
&= \alpha_{n}^{2} \Vert x_{n}-x^{*}-\beta(\nabla f(x_{n})-\nabla f(x^{*})) \Vert^{2} \nonumber\\
&\quad +(1-\alpha_{n})^{2} \Vert \hat{T}(\omega_{n}^{*},x_{n})-x^{*} \Vert^{2} \nonumber\\
&\quad +2 \alpha_{n}(1-\alpha_{n}) \langle x_{n}-x^{*} \nonumber\\
&\quad -\beta (\nabla f(x_{n})- \nabla f(x^{*})),\hat{T}(\omega_{n}^{*},x_{n})-x^{*} \rangle \nonumber\\
&\quad -2 \alpha_{n}  \langle \beta \nabla f(x^{*}),x_{n+1}-x^{*} \rangle - \alpha_{n}^{2}  \Vert \beta \nabla f(x^{*}) \Vert^{2}. 
\end{align*}
From (\ref{10}), quasi-nonexpansivity property of $\hat{T}(\omega^{*},x)$, and Cauchy--Schwarz inequality, we obtain
$$\langle x_{n}-x^{*}-\beta (\nabla f(x_{n})- \nabla f(x^{*})),\hat{T}(\omega_{n}^{*},x_{n})-x^{*} \rangle$$
\begin{equation}\label{caushy}
\leq (1-\gamma) \Vert x_{n}-x^{*} \Vert^{2}.
\end{equation}
From (\ref{10}), we also obtain
\begin{equation}\label{pprrov1}
\Vert x_{n}-x^{*}-\beta(\nabla f(x_{n})-\nabla f(x^{*})) \Vert^{2} \leq (1-\gamma)^{2} \Vert x_{n}-x^{*} \Vert^{2}.
\end{equation}
Therefore, from (\ref{caushy}), (\ref{pprrov1}), and quasi-nonexpansivity property of $\hat{T}(\omega^{*},x)$, we have
\begin{align*}
&\quad \Vert x_{n+1}-x^{*} \Vert^{2} \nonumber\\
&=\alpha_{n}^{2} \Vert x_{n}-x^{*}-\beta(\nabla f(x_{n})-\nabla f(x^{*})) \Vert^{2} \nonumber\\
&\quad +(1-\alpha_{n})^{2} \Vert \hat{T}(\omega_{n}^{*},x_{n})-x^{*} \Vert^{2} \nonumber\\
&\quad +2 \alpha_{n}(1-\alpha_{n}) \langle x_{n}-x^{*}-\beta (\nabla f(x_{n})- \nabla f(x^{*})), \nonumber\\
&\quad \hat{T}(\omega_{n}^{*},x_{n})-x^{*} \rangle \nonumber\\
&\quad -2 \alpha_{n}  \langle \beta \nabla f(x^{*}),x_{n+1}-x^{*} \rangle - \alpha_{n}^{2}  \Vert \beta \nabla f(x^{*}) \Vert^{2} \nonumber\\
&\leq (1-2\gamma \alpha_{n}) \Vert x_{n}-x^{*} \Vert^{2} \nonumber\\
&\quad +\alpha_{n} (\gamma^{2} \alpha_{n} \Vert x_{n}-x^{*} \Vert^{2}-2  \langle \beta \nabla f(x^{*}),x_{n+1}-x^{*} \rangle) \nonumber\\
&=(1-\gamma \alpha_{n}) \Vert x_{n}-x^{*} \Vert^{2}-\gamma \alpha_{n} \Vert x_{n}-x^{*} \Vert^{2} \nonumber\\
&\quad +\alpha_{n}(\gamma^{2} \alpha_{n} \Vert x_{n}-x^{*} \Vert^{2}-2  \langle \beta \nabla f(x^{*}),x_{n+1}-x^{*} \rangle).  
\end{align*}
Since $\gamma \alpha_{n} \Vert x_{n}-x^{*} \Vert^{2} \geq 0$, we have
\begin{align*}
&\quad (1-\gamma \alpha_{n}) \Vert x_{n}-x^{*} \Vert^{2}-\gamma \alpha_{n} \Vert x_{n}-x^{*} \Vert^{2} \nonumber\\
&\quad +\alpha_{n}(\gamma^{2} \alpha_{n} \Vert x_{n}-x^{*} \Vert^{2}-2  \langle \beta \nabla f(x^{*}),x_{n+1}-x^{*} \rangle) \nonumber\\
&\leq( 1-\gamma \alpha_{n}) \Vert x_{n}-x^{*} \Vert^{2} \nonumber\\
&\quad +\alpha_{n}(\gamma^{2} \alpha_{n} \Vert x_{n}-x^{*} \Vert^{2}-2  \langle \beta \nabla f(x^{*}),x_{n+1}-x^{*} \rangle) 
\end{align*}
or, finally, 
\begin{align}
&\Vert x_{n+1}-x^{*} \Vert^{2} \leq (1-\gamma \alpha_{n}) \Vert x_{n}-x^{*} \Vert^{2}+   \nonumber \\
&\quad \gamma \alpha_{n} (\frac{\gamma^{2} \alpha_{n} \Vert x_{n}-x^{*} \Vert^{2}-2  \langle \beta \nabla f(x^{*}),x_{n+1}-x^{*} \rangle}{\gamma}).  \label{llppp}
\end{align}
From \textit{Step 1}, \textit{Step 2}, (\ref{optimality}), and Theorem 1 \textit{(a)}, we obtain 
\begin{align}
&\displaystyle \lim_{n \longrightarrow \infty} (\gamma^{2} \alpha_{n} \Vert x_{n}-x^{*} \Vert^{2}-2 \beta \langle \nabla f(x^{*}),x_{n+1}-x^{*} \rangle)   \nonumber \\
&\leq 0 \quad{\textit{almost surely}}.  \label{lasssstt}
\end{align}
According to Lemma 1 by setting 
\begin{align*}
&a_{n}=\Vert x_{n}-x^{*} \Vert^{2}, \\
&b_{n}=\gamma \alpha_{n}, \\
&h_{n}=(\frac{\gamma^{2} \alpha_{n} \Vert x_{n}-x^{*} \Vert^{2}-2 \beta \langle \nabla f(x^{*}),x_{n+1}-x^{*} \rangle}{\gamma}),
\end{align*}
we obtain from (\ref{llppp}), (\ref{lasssstt}), and Theorem 1 \textit{(b)} that 
$$\displaystyle \lim_{n \longrightarrow \infty} \Vert x_{n}-x^{*} \Vert^{2}=0 \quad{\textit{almost surely}}.$$ Therefore, $\{ x_{n} \}_{n=0}^{\infty}$ converges almost surely to $x^{*}$. Thus the proof of Theorem 1 is complete. 

Since almost sure convergence in general does not imply mean square convergence and vice versa, we show the mean square convergence of the random sequence generated by Algorithm (\ref{7}) in the following theorem.

\textbf{Theorem 2:} Consider Problem 1 with Assumptions 1 and 2. Suppose that $\beta \in (0,\frac{2 \rho}{K^{2}})$ and $\alpha_{n} \in [0,1], n \in \mathbb{N} \cup \lbrace 0 \rbrace$, satisfies \textit{(a)} and \textit{(b)} in Theorem 1. Then starting from any initial point, the sequence generated by (\ref{7}) globally converges in mean square to the unique solution of the problem. 

\textit{Proof:} From \textit{Step 1}, Theorem 1, and Lemma 2, one can prove Theorem 2.

\section{Conclusions}

In this paper, we have defined a new optimization framework, i.e., minimization of a convex function over fixed value point set of a quasi-nonexpansive random operators on Hilbert spaces. This optimization includes the optimization defined in \cite{alavianiACC17}-\cite{alavianiTAC} as a special case. We have shown that the proposed algorithm in \cite{alavianiTAC} can converge almost surely and in mean square to the global solution of this optimization problem under suitable assumptions. 

\appendices
\section{}

\textit{Proof of Lemma 3:}

\textit{(i)} The proof is the same as the proof of part \textit{(i)} of Lemma 5 in \cite{alavianiTAC}.

\textit{(ii)}  

We have from quasi-nonexpansivity of $T(\omega^{*},x)$ for arbitrary $x \in \mathcal{H}$ that
\begin{equation}\label{ii1}
\Vert T(\omega^{*},x)-z \Vert^{2} \leq \Vert x-z \Vert^{2}, \forall z \in FVP(T),\forall \omega^{*} \in \Omega^{*}.
\end{equation}
From (\ref{hilbertinequ}), we obtain for all $z \in FVP(T)$ and for all $\omega^{*} \in \Omega^{*}$ that
\begin{align}
\Vert T(\omega^{*},x)-z \Vert^{2} &=\Vert T(\omega^{*},x)-x+x-z \Vert^{2} \nonumber\\
&=\Vert T(\omega^{*},x)-x \Vert^{2}+\Vert x-z \Vert^{2} \nonumber\\
&\quad +2 \langle T(\omega^{*},x)-x,x-z \rangle.\label{ii2}                          
\end{align}
Substituting (\ref{ii2}) for (\ref{ii1}) yields
\begin{equation}\label{ii3}
2 \langle x-T(\omega^{*},x),x-z \rangle \geq \Vert T(\omega^{*},x)-x \Vert^{2}. 
\end{equation}
From the definition of $\hat{T}(\omega_{n}^{*},x_{n})$ (see (\ref{7})), substituting $x-T(\omega^{*},x)=\frac{x-\hat{T}(\omega^{*},x)}{\eta}$ for the left hand side of the inequality (\ref{ii3}) implies \textit{(ii)}. Thus the proof of part \textit{(ii)} of Lemma 3 is complete.

\textit{(iii)}

We have from quasi-nonexpansivity of $T(\omega^{*},x)$ for $z \in FVP(T)$ and arbitrary $x \in \mathcal{H}$ that
\begin{align*}
&\quad \Vert \hat{T}(\omega^{*},x)-z \Vert  \\
&\leq (1-\eta) \Vert x-z \Vert+\eta \Vert T(\omega^{*},x)-z \Vert \\
&\leq (1-\eta) \Vert x-z \Vert+\eta \Vert x-z \Vert \\
&= \Vert x-z \Vert, \forall \omega^{*} \in \Omega^{*}.
\end{align*}
Therefore, $\hat{T}(\omega^{*},x)$ is a quasi-nonexpansive random operator, and the proof of part \textit{(iii)} of Lemma 3 is complete.


\begin{thebibliography}{00}
	
	
	
	\bibitem {alavianiACC17} S. Sh. Alaviani and N. Elia, Distributed multi-agent convex optimization over random digraphs, \textit{Proc. of Amer. Cont. Conf.}, Sheraton Seattle Hotel, May 24-26, Seattle, USA, pp. 5288--5293, 2017.
	
	\bibitem {alavianiTAC} ---------------, Distributed multi-agent convex optimization over random digraphs, \textit{IEEE Trans. on Automatic Control}, to appear
	
	
	\bibitem {signal} P. L. Combettes, A block-iterative surrogate constraint splitting method for quadratic signal recovery, \textit{IEEE Trans. on Signal Processing}, vol. 51, pp. 1771--1782, 2003.
	
	
	
		\bibitem {dataregression} S. S. Ram, A. Nedi\'{c}, and V. V. Veeravalli, A new class of distributed optimization algorithms: application to regression of distributed data, \textit{Optim. Methods Softw.}, vol. 27, pp. 71--88, 2012.
	
	
	
		\bibitem {boyd} S. Boyd and L. Vandenberghe, \textit{Convex Optimization}, Cambridge University Press: New York, 2004. 
	
	
	
		\bibitem {nicola1} J. Wang and N. Elia, Control approach to distributed optimization, \textit{Proc. of $48^{th}$ Annual Allerton Conf.}, Alerton House, UIUC, Illinois, USA, Sep. 29-Oct. 1, pp. 557-561, 2010.

	
	\bibitem {nicola2} ---------------, A control perspective for centralized and distributed convex optimization, \textit{Proc. of $50^{th}$ IEEE Conf. on Decision and Control and European Control Conf.}, Dec. 12-15, Orlando, FL, USA, pp. 3800-3805, 2011.
	
	
	\bibitem {shamma} J. R. Marden, G. Arslan, and J. S. Shamma, Cooperative control and potential games, \textit{IEEE Trans. on Systems, Man, and Cybernetics-part B: Cybernetics}, vol. 39, pp. 1393--1407, 2009.
	
	\bibitem {nali} N. Li and J. R. Marden, Designing games for distributed optimization, \textit{IEEE J. Selec. Topic. Sign. Process.}, vol. 7, pp. 230--242, 2013.
	
	\bibitem {alavianiCDC} S. Sh. Alaviani and N. Elia, A distributed algorithm for solving linear algebraic equations over random networks, \textit{Proc. of IEEE Conf. on Decision and Control}, Dec. 17-19, Miami Beach, FL, USA, pp. 83--88, 2018.
	
	\bibitem {alavianiACC19} -----------------, Distributed average consensus over random networks, \textit{Proc. of Amer. Cont. Conf.}, July 10-12, Philadelphia, PA, USA, pp. 1854--1859, 2019.
	
	
	\bibitem {alavianidissertation} S. Sh. Alaviani, Applications of fixed point theory to distributed optimization, robust convex optimization, and stability of stochastic systems, Ph.D. dissertation, Dep. Elec. Comput. Eng., Iowa State University, Ames, Iowa, USA, 2019.
	
	
	
	
	\bibitem {quasinonexpansive} W. G. Dotson, Jr., Fixed points of quasi-nonexpansive mappings, \textit{J. Austral. Math. Soc.}, vol. 13, pp. 167--170, 1972.
	
	
	
		\bibitem {ssdd} H. K. Xu, ÒIterative algorithms for nonlinear operators,Ó \textit{J. London Math. Soc.}, vol. 66, pp. 240-256, 2002.
	
	 
	
	
	
	
	
	
	
	
	
	
	
	
	
	
	
	
	
	
	
	
	
	
	
	
	
	
	
	
	
	
\end{thebibliography}
\end{document}